\documentstyle[12pt,titlepage]{ioplppt}
\begin{document}
\jl{1}
\title{Car accidents and number of stopped cars due to road blockage
on a one-lane highway}[Car accidents and blockage]
\author{N Boccara\dag\ddag, H Fuk\'s\dag\ and Q Zeng\dag}
\address{\dag University of Illinois at Chicago, Dept. of Physics, Chicago,
IL 60607-7059, USA}
\address{\ddag DRECAM/SPEC, CE Saclay, 91191 Gif sur Yvette, France}
\maketitle
\begin{abstract}
Within the framework of a simple model of car traffic on a one-lane
highway, we study the probability for car accidents to occur when
drivers do not respect the safety distance between cars, and, as a
result of the blockage during the time $T$ necessary to clear the
road, we determine the number of stopped cars as a function of car
density. We give a simple theory in good agreement with our numerical
simulations.
\end{abstract}
\section{Car accidents}
In the past few years, many highway traffic models formulated in terms
of cellular automata have been studied, both in one [1--5]
\nocite{Nagel92,Nagel93,Schadschneider93,Schreckenberg95,Vilar94}
and two dimensions [6--7] \nocite{Biham92,Tadaki94,Tadaki95,Molera95}.
For the one-dimensional case, several topological variations of the
basic model have been proposed, including road crossing
\cite{Ishibashi96}, road with junction \cite{Benjamin96} and two-lane
highway [12--14]\nocite{Nagatani94c,Nagatani96a,Rickert96}. Recently,
experimental features and complex spatiotemporal structures of real
traffic flows have been investigated \cite{Kerner96c,Kerner96a}.

One of the simplest models is defined on a one-dimensional lattice of
$L$ sites with periodic boundary conditions. Each site is either
occupied by a vehicle, or empty. The velocity of each vehicle is an
integer between 0 and $v_{\rm max}$.  If $x(i,t)$ denotes the position
of the $i$th car at time $t$, the position of the next car ahead at
time $t$ is $x(i+1,t)$. With this notation, the system evolves
according to a synchronous rule given by
\begin{equation}
x(i,t+1)=x(i,t)+v(i,t+1),
\label{eq1}
\end{equation}
where
\begin{equation}
 v(i,t+1)=\min\big(x(i+1,t)-x(i,t)-1,x(i,t)-x(i,t-1)+a,v_{\rm max}\big)
 \label{eq2}
\end{equation}
is the velocity of car $i$ at time $t+1$. $x(i+1,t)-x(i,t)-1$ is the
gap (number of empty sites) between cars $i$ and $i+1$ at time $t$,
$x(i,t)-x(i,t-1)$ is the velocity $v(i,t)$ of car $i$ at time $t$, and
$a$ is the acceleration. $a=1$ corresponds to the deterministic model
of Nagel and Schreckenberg \cite{Nagel92}, while the case $a=v_{\rm
max}$ has been considered by Fukui and Ishibashi \cite{Fukui96c}. In
this last case, the evolution rule can be written
\begin{equation}
 x(i,t+1)=x(i,t)+\min\big(x(i+1,t)-x(i,t)-1,v_{\rm max}\big).
\end{equation}
This is a cellular automaton rule whose radius is equal to $v_{\rm
max}$. The case $a<v_{\rm max}$ is a second order rule, that is, the
state at time $t+1$ depends upon the states at times $t$ and $t-1$.

We studied the probability for a car accident to occur when drivers do
not respect the safety distance. More precisely, if at time $t$, the
velocity $v(i+1,t)$ of car $i+1$ was positive, expecting this velocity
to remain positive at time $t+1$, the driver of car $i$ increases the
safety velocity $v(i,t+1)$ given by (\ref{eq2}) by one unit, with a
probability $p$. The evolution rule (\ref{eq1}) is then replaced by
\begin{equation}
 \hbox{if}\quad v(i+1,t)>0,\quad\hbox{then}\quad
 x(i,t+1)=x(i,t)+v(i,t+1)+\Delta V,
\end{equation}
where $\Delta V$ is a Bernoulli random variable which takes the value
1 with probability $p$ and zero with probability $1-p$. If
$v(i+1,t+1)=0$, it is clear that this careless driving will result in
an accident.

When the car density $\rho$ is less than the critical car density
$\rho_c=(1+v_{\rm max})^{-1}$, the average number of empty sites
between two consecutive cars is larger than $v_{\rm max}$, the
fraction $n_0$ of stopped cars is zero and no accident can occur. If
$\rho>\rho_c$, the average velocity is less than $v_{\rm max}$, $n_0$
increases with $\rho$ and careless driving will result in a number of
accidents. This number will, however, go to zero for $\rho=1$, since,
in this case, all cars are stopped. The probability for a car accident
to occur should, therefore, reach a maximum for a car density $\rho$
between $\rho_c$ and 1.

Neglecting time correlations, we may determine an approximate
probability for an accident to occur. Let $n$ be the number of empty
sites between cars $i$ and $i+1$ at time $t$. If the three conditions
\begin{equation}
 0\le n\le v_{\rm max},\quad v(i+1,t)>0,\quad v(i+1,t+1)=0
\end{equation}
are satisfied then car $i$ will cause an accident at time $t+1$, with
a probability $p$. Therefore, the value $P_{as}$ of the probability
{\it per site\/} and {\it per time step\/} for an accident to occur is
given by
\begin{equation}
P_{as}=pn_0(1-n_0)\sum_{n=0}^{v_{\rm max}}\rho^2(1-\rho)^n=
p\rho\big(1-(1-\rho)^{v_{\rm max+1}}\big)n_0(1-n_0).
\end{equation}
Dividing by $\rho$, one obtains the probability {\it per car} and {\it per
time step\/} for an accident to occur
\begin{equation}
P_{ac}=p\rho\big(1-(1-\rho)^{v_{\rm max+1}}\big)n_0(1-n_0).
\label{eq7}
\end{equation}

\begin{figure}[p]
\begin{center}
\input{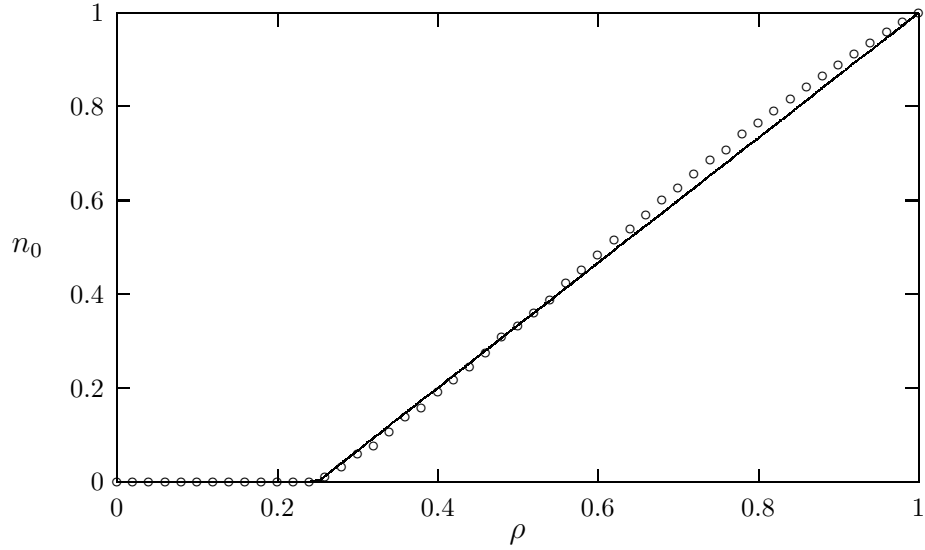}
\end{center}
\caption{Fraction of stopped cars $n_0$ as a function of the car density
$\rho$ for $v_{\rm max}=3$ and $a=1$. The solid line represents the linear
approximation $n_0=(\rho-\rho_c)/(1-\rho_c)$.}
\label{fig1}
\end{figure}

\begin{figure}[p]
\begin{center}
\input{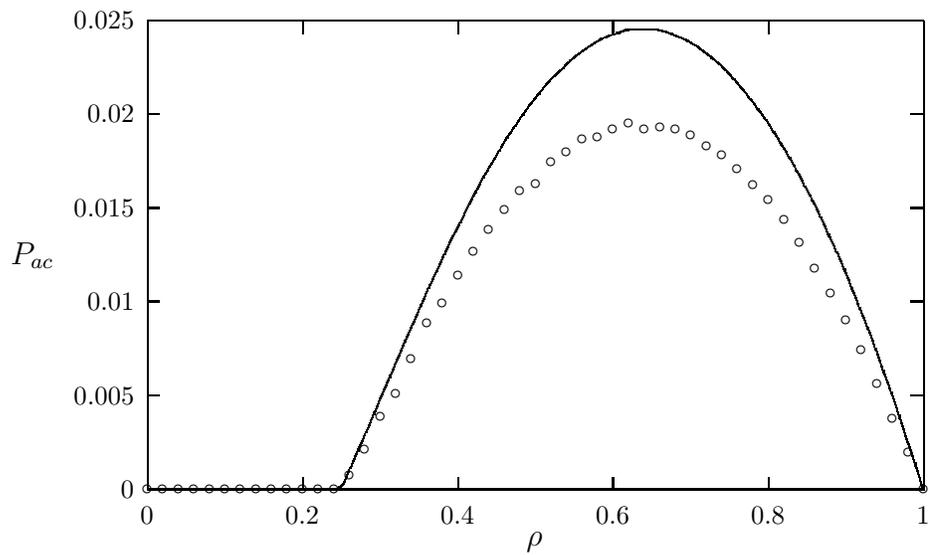}
\label{fig2}
\end{center}
\caption{Probability $P_{ac}$ per car and per time step for an accident to
occur as a function of the car density $\rho$ for $v_{\rm max}=3$ and
$a=1$. The solid line corresponds to the approximation given by
(\protect{\ref{eq9}}).}
\end{figure}

The simplest approximate expression for the fraction $n_0$ of stopped
cars as a function of car density, satisfying the above-mentioned
conditions, is
\begin{equation}
n_0=\frac{\rho-\rho_c}{1-\rho_c}.
\label{eq8}
\end{equation}
This approximation is rather crude. In particular, it neglects the
fact that $n_0$ should depend upon $v_{\rm max}$. However, as shown in
Figure~\ref{fig1}, for $v_{\rm max}=3$ and $a=1$, this linear
approximation is in rather good agreement with our numerical results.
Substituting (\ref{eq8}) in (\ref{eq7}) yields
\begin{equation}
P_{ac}=p\rho\big(1-(1-\rho)^{v_{\rm max+1}}\big)
\frac{(\rho-\rho_c)(1-\rho)}{(1-\rho_c)^2}.
\label{eq9}
\end{equation}
Figure~\ref{fig2} represents the probability $P_{ac}$ as a function of
$\rho$ determined numerically and its value given by (\ref{eq9}).

\section{Stopped cars due to blockage}
As a result of an accident (or any other cause such as road works),
the traffic is blocked during the time $T$ necessary to clear the
road. For a given $T$, the number $N(\rho)$ of blocked cars is clearly
an increasing function of the car density $\rho$. To determine the
expression of $N$, we shall distinguish two regimes.

If $\rho\le\rho_c$, the average car velocity is $v_{\rm max}$. Since
the average number of empty sites between two consecutive cars is
equal to
\begin{equation}
d(\rho)=\frac{(1-\rho)}{\rho},
\end{equation}
the line of stopped cars increases by one unit during the time interval
\begin{equation}
\frac{v_{\rm max}}{d(\rho)}.
\end{equation}
Hence, during the time $T$ the number $N$ of blocked cars is given by
\begin{equation}
N(\rho)=T\frac{\rho(1-\rho_c)}{(1-\rho)\rho_c}.
\label{eq12}
\end{equation}

\begin{figure}[p]
\begin{center}
\input{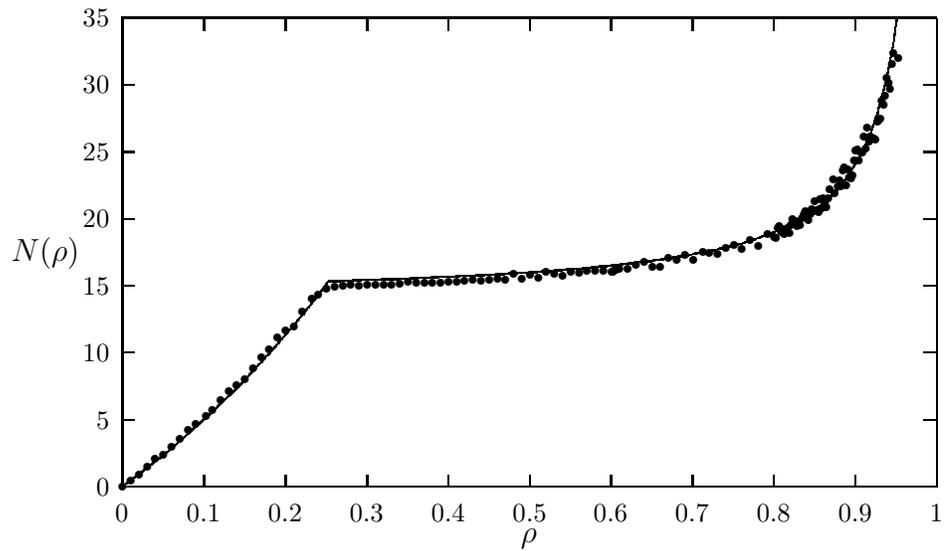}
\end{center}
\caption{Average number of blocked cars as a function of the car
density $\rho$. The solid line corresponds to the approximate
expressions (\protect{\ref{eq12}}) and (\protect{\ref{eq13})}.}
\label{fig3}
\end{figure}

For $\rho=\frac{1}{2}$ the average number of empty sites between two
consecutive cars is equal to $d(\frac{1}{2})=1$, and the average
number of blocked cars increases by one unit at each time step.
$N(\rho)$ increases, therefore, from $T$ to $T+1$ when $\rho$
increases from $\rho_c$ to $\frac{1}{2}$ (note that the first blocked
site is occupied). When $\rho>\frac{1}{2}$, $d(\rho)<1$, and the
average number of blocked cars is then given by
\begin{equation}
N(\rho)=T+\frac{1}{d(\rho)}=T+\frac{\rho}{1-\rho}.
\label{eq13}
\end{equation}
Figure~\ref{fig3} represents the average number of blocked cars as a
function of the car density $\rho$. The agreement with the approximate
expressions of $N(\rho)$, given by (\ref{eq12}) and (\ref{eq13}), is
very good.


\end{document}